# Magnetically tunable optical bound states in the continuum with arbitrary polarization and intrinsic chirality

Qing-an Tu[1, #], Hongxin Zhou[1, #], Yan Meng[2, *], Maohua Gong[1, †], Zhen Gao[1, ‡]

[1]*State Key Laboratory of Optical Fiber and Cable Manufacturing Technology, Department of Electronic and Electrical Engineering, Guangdong Key Laboratory of Integrated Optoelectronics Intellisense, Southern University of Science and Technology, Shenzhen 518055, China*

[2]*School of Electrical Engineering and Intelligentization, Dongguan University of Technology, Dongguan, 523808, China*

**Abstract**

Optical bound states in the continuum (BICs), which are exotic localized eigenstates embedded in the continuum spectrum and topological polarization singularity in momentum space, have attracted great attentions in both fundamental and applied physics. Here, based on magneto-optical photonic crystal slab placed in external magnetic fields to break the time-reversal symmetry, we theoretically demonstrate magnetically tunable BICs with arbitrary polarization covering the entire Poincaré sphere and efficient off-Γ chiral emission of circularly polarized states. More interestingly, by further breaking the in-plane inversion symmetry of the magneto-optical photonic crystal slab to generate a pair of circularly polarized states (*C* point) spawning from the eliminated BICs and tuning the external magnetic field strength to move one *C* point to the Γ point, one at-Γ intrinsic chiral BICs with near-unity circular dichroism exceeding 0.99 and a high quality factor of 46000 owing to the preserved out-of-plane mirror symmetry can be observed. These findings may lead to a plethora of potential applications in chiral-optical effects, structured light, and tunable optical devices.

## 1. Introduction

Recently, optical bound states in the continuum (BICs)[1-3] have played a central role in topological polarization manipulation[4-6], chiral light-matter interaction[7-9], chiral emission[10-12] and ultrahigh quality factor (*Q* factor) optical resonators[13-15]. However, since the topological properties of BICs are mainly determined by the geometrical symmetry of the photonic structures, it is notoriously difficult to manipulate their properties without modifying the photonic structure. For instance, by breaking the in-plane inversion ($C_2$) symmetry[16] or the $C_6$ symmetry[17] of the photonic crystal (PhC) slab, pairs of circularly polarized states (*C* points) with half topological charge can spawn from the eliminated BICs manifesting as a momentum-space vortex polarization singularity (*V* point). While for a PhC slab with inversion symmetry but broken out-of-plane mirror ($\sigma_z$) symmetry[18,19], the annihilation of two *C* points of downward radiation could generate exotic unidirectional guided resonances that radiate only in the upward direction. More interestingly, by breaking the $C_2$ and $\sigma_z$ symmetries simultaneously, arbitrarily polarized BICs in bilayer-twisted PhC slabs[20] and chiral BICs in dielectric metasurfaces[21-24] can be realized. Nonetheless, once these photonic structures are fabricated and settled, it is rather difficult to dynamically tune the topological properties of BICs without modifying the geometrical structures, thus seriously limiting their tunability and practical applications.

On the other hand, magneto-optical (MO) modulation of light provides a new strategy to dynamically manipulate the topological properties of BICs without modifying the photonic structure[25-27]. For example, both the light intensity[28] and topological polarization singularities in momentum space[29] can be modulated by the external magnetic field in one-dimensional MO grating with Faraday configuration. Besides, extreme nonreciprocity and near-unity magnetic circular dichroism can be achieved in subwavelength magnetic metasurface supporting BICs[30]. These findings significantly promote the applications of MO effect to modulate the topological properties of BICs actively. However, the magnetically tunable BICs with arbitrary polarization covering the whole Poincaré sphere and intrinsic chirality with near-unity circular dichroism and ultrahigh *Q* factor remain elusive so far.

In this work, based on two-dimensional (2D) MO PhC slabs placed in external magnetic fields with time-reversal symmetry (TRS) breaking, we theoretically propose a new paradigm for realizing magnetically tunable BICs with arbitrary polarization, off-Γ chiral emission, and at-Γ intrinsic chirality. By continuously modulating the strength of the external magnetic field, we find that the far-field polarization around BICs can be dynamically tuned from linear to elliptical and eventually to circular polarization, achieving full coverage of arbitrary polarization on the entire Poincaré sphere. Moreover, under a proper external magnetic field, we observe off-Γ chiral emission of circularly polarized light under right-circularly polarized (RCP) and left-circularly polarized (LCP) incidences. Interestingly, by further breaking the $C_2$ symmetry, a pair of *C* points can spawn from the eliminated BICs. Significantly, by tuning the external magnetic field strength, we can gradually move one *C* point to the Γ point, achieving an intrinsic chiral BICs with near-unity circular dichroism exceeding 0.99 and an ultrahigh *Q* factor up to 46000 while preserving the $\sigma_z$ symmetry. Our findings open new avenues for dynamically tuning the topological properties of BICs without modifying the photonic structures and may lead to promising applications in polarization manipulation[31-33], chiral light source and detector[34-36], and magnetic tunable photonic devices[37,38].

## 2. Results and Discussion

To realize magnetically tunable BICs with arbitrary polarization and intrinsic chirality, we start with a free-standing 2D MO PhC slab consisting of a square array of circular air holes exhibiting $C_{4v}$ symmetry, as shown in Fig. 1(a). The inset shows the unit cell of the MO PhC slab with lattice constant $a$ = 1.4 μm, thickness $h$ = 1.2 μm, and air hole diameter $d = 1$ μm. Under perpendicular external magnetic field (along the $z$ direction) to break the TRS, the optical response of the MO PhC slab can be expressed by the following permittivity tensor[29]:

$$\vec{\varepsilon} = \begin{pmatrix} \varepsilon & i\delta & 0 \\ -i\delta & \varepsilon & 0 \\ 0 & 0 & \varepsilon \end{pmatrix}, \tag{1}$$

where $\varepsilon$ represents the dielectric constant which is approximately 4 within the frequency regime of interest, and $\delta$ describes the magnetization-induced gyration of the material which is proportional to the external magnetic field strength. The governing equation of the MO PhC slab is given as[39]:

$$\nabla \times (\nabla \times \boldsymbol{E}) - \omega^2 \mu_0 \vec{\varepsilon} \cdot \boldsymbol{E} = 0, \tag{2}$$

where $\omega$ is the angular frequency and $\mu_0$ is the magnetic permeability. By substituting the permittivity in Eq. (1) into Eq. (2), two characteristic solutions can be achieved (see more details in Supplementary Information Note S1):

$$E_z = 0, E_x = \frac{k_x^2 - \omega^2 \mu_0 \varepsilon}{k_x k_y - i\omega^2 \mu_0 \delta} E_y, \beta_1^2 = \frac{\omega^2 \mu_0 (\varepsilon^2 - \delta^2)}{\varepsilon}, \tag{3}$$

$$E_z \neq 0, E_x = E_y = 0, \beta_2^2 = \omega^2 \mu_0 \varepsilon, \tag{4}$$

the characteristic solutions in Eq. (3) and Eq. (4) correspond to the transverse electric (TE) and transverse magnetic (TM) modes, respectively. The third term in Eq. (3) represents the propagation constant as a function of $\omega$ and $\delta$, while the one in Eq. (4) is independent of $\delta$. It can be observed that the two in-plane components of the electric field ($E_x$ and $E_y$) in the solution of TE modes are related by a function of $\delta$, implying that only the polarization of the TE mode can respond to the external magnetic field. Thus, in this work we consider only the TE mode to investigate the magnetically tunable BICs.

The band structure of the MO PhC slab with zero external magnetic field ($\delta = 0$) is shown in Fig. 1(b), in which the lower band of interest is highlighted in red. The inset shows the field profile of the TE-like eigenmode ($|H_z|$) which exhibits a typical electric quadrupole mode. Owning to the $C_2$ and $\sigma_z$ symmetries of the MO PhC slab, a symmetry-protected BICs can be observed at the Γ point[1]. To confirm the existence of the symmetry-protected BICs, the far-field polarization indicated by the azimuthal angle ($\psi$) and ellipticity angle ($\chi$) is extracted from the Stokes parameters[17] and shown in the lower panel of Fig. 1(c) (see more details in Supplementary Information Note S2). The symmetry-protected BICs at the Γ point is surrounded by far-field linear polarization indicated by short blue lines. The topological charge ($q$) of the symmetry-protected BICs can be calculated from the azimuthal angle of the far-field polarization[4]:

$$q = \frac{1}{2\pi} \oint_L d\mathbf{k}_\parallel \cdot \nabla_{\mathbf{k}_\parallel} \psi(\mathbf{k}_\parallel), \tag{5}$$

where $\psi(\mathbf{k}_\parallel)$ represents the azimuthal angle between the long axis of polarization and the $x$-axis, and $L$ is a closed path (counterclockwise direction) enclosing the target point in the momentum space. Therefore, the topological charge of BICs can be obtained by winding the linear polarization around the Γ point ($q = -1$). We then calculate the $Q$ factor of the eigenstates around the Γ point. As shown in Fig. 1(d), an ultrahigh $Q$ factor (>$10^9$) can be observed at the Γ point. Both the nonzero topological charge and ultrahigh $Q$ factor confirm the existence of the symmetry-protected BICs at the Γ point. Interestingly, when we continue to increase the external magnetic field strength, we find that the far-field polarization around the symmetry-protected BICs can evolve gradually from linear (lower panel) to elliptical (middle panel) and eventually to circular polarization (upper panel), as shown in Fig. 1(c), indicating that magnetically tunable BICs with arbitrary polarization covering the entire Poincaré sphere can be achieved by tuning the external magnetic field strength.

To quantitatively characterize the tunability of the external magnetic field to the symmetry-protected BICs, we investigate the far-field polarization around the BICs under different values of $\delta$ (external magnetic field strength) in Figs. 2(a)-2(d). As $\delta$ gradually increases from 0 to 0.3, the far-field polarization around BICs changes from linear to elliptical and finally to circular polarization. Figs. 2(e)-2(h) show the far-field ellipticity under the corresponding external magnetic fields. As the external magnetic field strength increases, the ellipticity of the far-field polarization gradually evolves from 0 to 1, agreeing well with the far-field polarization shown in Figs. 2(a)-2(d) and further verifying the magnetically tunable BICs with arbitrary polarization covering the entire Poincaré sphere.

Next, we calculate the Stokes parameters around an iso-frequency contour with a radius of $0.03\pi/a$ (red dashed circle in Figs. 2(a)-2(d)) and map them onto the Poincaré sphere as red rings, as shown in Fig. 2(i). As $\delta$ increases, the red circles and rings become darker for clarity. For $\delta = 0$, the red ring is located at the equator of the Poincaré sphere, indicating linear polarization states. When $\delta$ increases from 0 to 0.3, the red ring gradually shrinks into a point close to the North Pole corresponding to RCP states. The red rings for $\delta = 0.1$ and $\delta = 0.2$ are located between the equator and the North Pole, corresponding to elliptically polarized states. The latitude of the red ring gradually varies from 0° to 90° on the Poincaré sphere with increasing magnetic field (the lower hemisphere can be covered by inverting the external magnetic field; see more details in Supplementary Information Fig. S1), indicating that the far-field polarization of the magnetically tunable BICs can fully cover the entire Poincaré sphere.

The arbitrary polarization tunability of the MO PhC slab can be explained by the MO effect. For instance, an incident wave with linear polarization can be converted to circular polarization emission and vice versa. The ellipticity of the far-field polarization can be derived[40,41]:

$$\rho = \tan\left(\frac{1}{2}\arcsin\left(\frac{2\,\mathrm{Im}(\eta)}{1+|\eta|^2}\right)\right), \tag{6}$$

where $\eta = \langle \tilde{E}_y \rangle / \langle \tilde{E}_x \rangle$, $\tilde{E}_x$ and $\tilde{E}_y$ are the projection components of the $s(p)$-polarization on the $x$-$y$ plane. The bracket indicates the time average of electric fields. According to Eq. (6), the evolution of ellipticity at the point of $k_x = 0.03\pi/a$ and $k_y = 0$ (triangles in Figs. 2(a)-2(d)) is shown in Fig. 2(j). As $\delta$ increases from 0 to 0.3, the ellipticity $\rho$ approximately evolves linearly from 0 to 1 with good agreement between the analysis (gray stars) and simulation (red line) results (see more details for the calculation of the far-field polarization in Supplementary Information Note S2). The excellent consistency between these two approaches indicates that the magnetically tunable arbitrary polarization indeed originates from the MO effect. Note that the existence of magnetically tunable BICs with arbitrary polarization is a universal phenomenon and can also be found in hexagonal MO PhC consisting of dielectric cylinders (see more details in Supplementary Information Note S4 and Fig. S2).

In addition to the magnetically tunable arbitrary far-field polarization, the MO effect also can induce TRS breaking and result in intriguing chirality phenomena such as chiral emission. Fig. 3(a) shows the circular polarization around Γ point in the momentum space with $\delta = 0.3$. The reflection spectra (color map) of the MO PhC slab illuminated by RCP and LCP waves are shown in Figs. 3(b) and 3(c), respectively. Here the in-plane wavevector $k_x$ ranges from $-0.05\pi/a$ to $0.05\pi/a$ and $k_y = 0$, corresponding to the red dashed line in Fig. 3(a). For both cases, it can be seen that the reflections are near zero at the Γ point because the eigenstates at the Γ point correspond to a symmetry-protected BICs with an infinite $Q$ factor and therefore are decoupled from the incident waves. When deviating from the Γ point, obvious reflection can be observed for the RCP incident wave in the frequency range from 152.5 THz to 153.5 THz, while the reflection is extremely low for the LCP incidence. The distinct reflection responses for incident waves with opposite circular polarization demonstrate the unique off-Γ chiral emission of the MO PhC slab. To explore the coverage of the chiral response in the momentum space, Figs. 3(e) and 3(f) present the reflection spectra of an iso-frequency contour at $f = 153$ THz (red dashed circle in Fig. 3(d)) for the RCP and LCP incidence, respectively, which exhibit chiral emission in the whole iso-frequency contour. Therefore, the MO PhC slab can exhibit chiral emission over a wide range of incident angles rather than only at a specific incident angle, providing a richer degree of freedom for chiral emission. Note that the transmission spectra exhibit similar chiral emission phenomena (see more details in Supplementary Information Fig. S3). Moreover, the MO PhC slabs possess stable working frequency and high $Q$ factor around the Γ point under different external magnetic field strengths (see more details in Supplementary Information Note S5 and Fig. S4).

To fully investigate the intriguing phenomena caused by the MO effect on the BICs, we further break the $C_2$ symmetry of the MO PhC slab by replacing the circular air holes in Fig. 1(a). In which the isosceles triangular air holes with side length $L_2 = 1$ μm and bottom length $L_1 = 0.92$ μm, while keeping the other geometric parameters unchanged, as shown in Fig. 4(a). When no external magnetic field is applied, the far-field polarization around the Γ point is shown in Fig. 4(b), in which the $V$ point (BICs at the Γ point) with an integer topological charge ($q = -1$) splits into a pair of $C$ points with half-integer topological charge ($q = -1/2$), governed by the conservation of total topological charge. The chirality of the right-hand and left-hand elliptical polarization states is indicated by red and blue colors, respectively. Due to the $\sigma_y$ symmetry of the configuration, these two *C* points are distributed symmetrically with respect to the $k_y$ axis.

Once a nonzero external magnetic field is applied to the MO PhC slab along the *z* direction, the two *C* points will move along the direction of positive $k_x$ axis (See Supplementary Information Fig. S6 for the complete evolution process). When the external magnetic field $\delta$ equals to 0.24, the *C* point with RCP (red dot) moves to the Γ point, indicating the emergence of an intrinsic chiral BICs, as shown in Fig. 4(c). Notably, the intrinsic chiral BICs at the Γ point with near-unity circular dichroism (CD) exceeding 0.99 (see more details in Supplementary Information Fig. S7). Fig. 4(d) shows the simulated azimuthal angles of the far-field polarization with respect to the major axis, in which we can see that the total winding angle of the polarization is π along counterclockwise loop enclosing the Γ point, indicating the topological charge of the polarization singularity at the Γ point is −1/2.

Note that previously reported intrinsic chiral BICs[24] break the $C_2$ and $\sigma_z$ symmetries simultaneously, while our design breaks the $C_2$ and TRS symmetries simultaneously. Generally, when both the $C_2$ and $\sigma_z$ symmetries are broken, chiral emission can only be found on one side of the photonic structure, and the *Q* factor of the intrinsic chiral BICs is limited due to the lower structural symmetry. In stark contrast, our intrinsic chiral BICs exhibits chiral characteristics on both sides of the PhC slab with an ultrahigh *Q* factor up to 46000 at the Γ point owning to the preserved $\sigma_z$ symmetry, as shown in Fig. 4(e), significantly enhancing the light-matter interactions and providing new perspectives on the realization of chiral laser[23]. Figs. 4(f) and 4(g) show the simulated reflection spectra for the LCP and RCP incidence with $\delta$ = 0.24, respectively, from which we can see that there is a vanishing point (red dashed circle) in the reflection spectra at the Γ point for LCP incidence, whereas the RCP incident wave can excite the eigenstate at the Γ point, indicating the presence of a *C* point with right-handed chirality at the Γ point. Moreover, there exists an off-Γ vanishing point (blue dashed circle) in the reflection spectrum for the RCP incidence, corresponding to a *C* point with left-handed chirality at the off-Γ point, as shown in Fig. 4(g). The results in Figs. 4(f) and 4(g) show good agreement with the polarization distribution in Fig. 4(c), further confirming the existence of the intrinsic chiral BICs.

## 3. Conclusion

In conclusion, we have theoretically proposed magnetically tunable BICs with arbitrary polarization and intrinsic chirality in MO PhC slab for the first time. By tunning the external magnetic field strength applied to MO PhC slab, the far-field polarization around the BICs in momentum space can gradually evolve from linear to circular, effectively covering the entire Poincaré sphere. In addition, off-Γ chiral emission can be achieved in a broad frequency range and with multiple incident angles for RCP and LCP incidence. More interestingly, by breaking the $C_2$ and TRS symmetries simultaneously, an intrinsic chiral BICs at the Γ point with near-unity circular dichroism and a high *Q* factor can be realized. Our work paves the way for exploring the magnetically tunable topological photonics of BICs with TRS breaking.

**Supporting Information**
Supporting Information is available from the Wiley Online Library or from the author.

**Acknowledgments**
Z.G. acknowledges the finding from the National Natural Science Foundation of China under grants No. 6231101016, 62375118, and 12104211, Guangdong Basic and Applied Basic Research Foundation under grant No.2024A1515012770, Shenzhen Science and Technology Innovation Commission under grants No. 20220815111105001 and 202308073000209, High level of special funds under grant No. G03034K004. Y. M acknowledges the support from the National Natural Science Foundation of China under Grant No. 12304484, Guangdong Basic and Applied Basic Research Foundation under grant No. 2024A1515011371.

**Conflict of Interest**
The authors declare no conflicts of interest.

**Data Availability Statement**
The data that support the findings of this study are available from the corresponding author upon reasonable request.

[#] Q. A. T. and H. X. Z contribute equally to this work.
* mengyan@dgut.edu.cn
[†] gongmh@sustech.edu.cn
[‡] gaoz@sustech.edu.cn

**Figures**

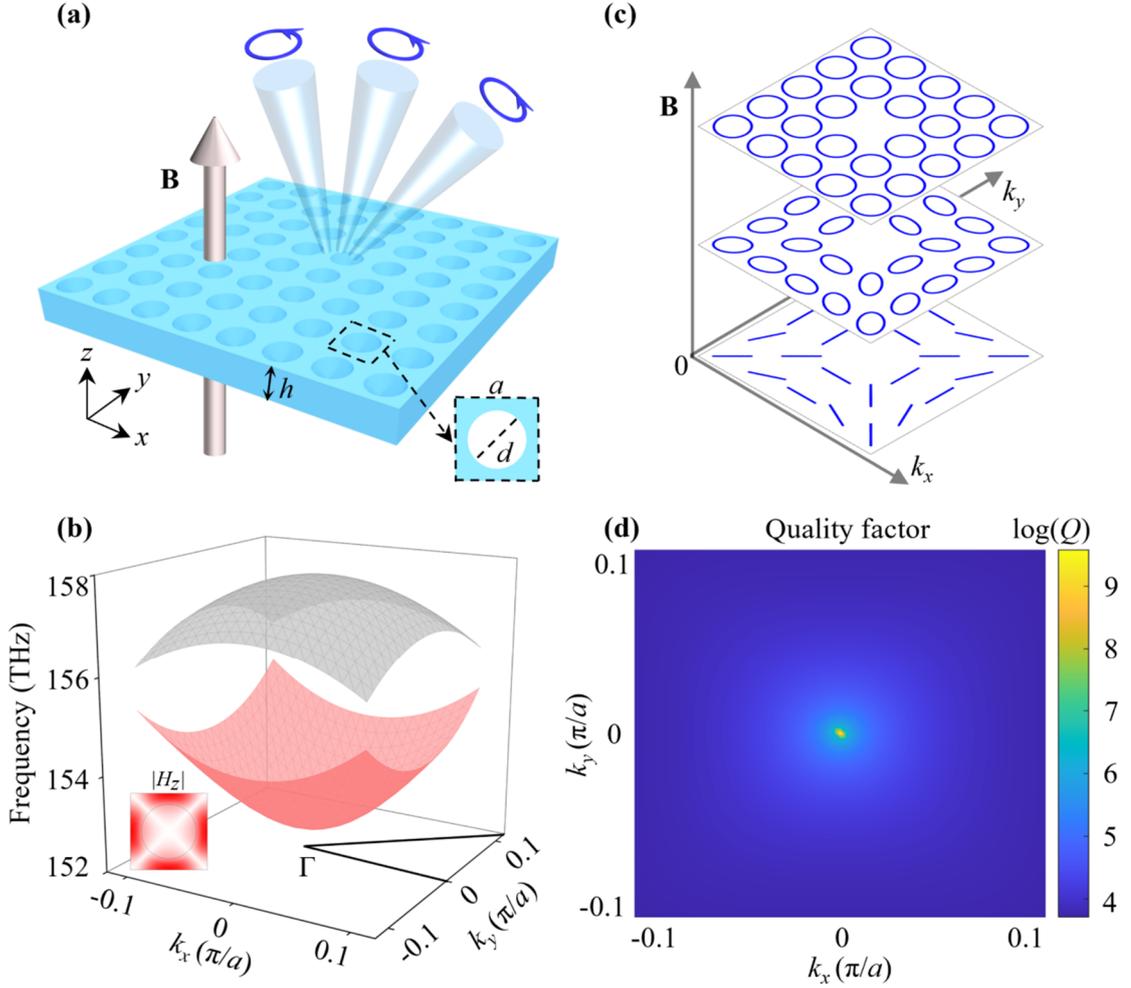

**Figure 1.** MO PhC slab supporting magnetically tunabel symmetry-protected BICs at the Γ point. (a) Schematic of a 2D MO PhC slab under external magnetic fields along the z direction. Arbitrary far-field polarization can be achieved by tuning the external magnetic fields. (b) Simulated band structure of the MO PhC slab with $\delta = 0$ (without external magnetic fields). The TE-like band of interest is highlighted in red. Inset shows the magnetic field profile of the eigenstate ($|H_z|$) at the Γ point. (c) The far-field polarization around the BICs evolves from linear (lower panel) to elliptical (middle panel) and finally to circular (upper panel) as the external magnetic field (**B**) increases. The topological polarization singularity at the Γ point indicates the existence of a symmetry-protected BICs. (d) Calculated $Q$ factor near the BICs.

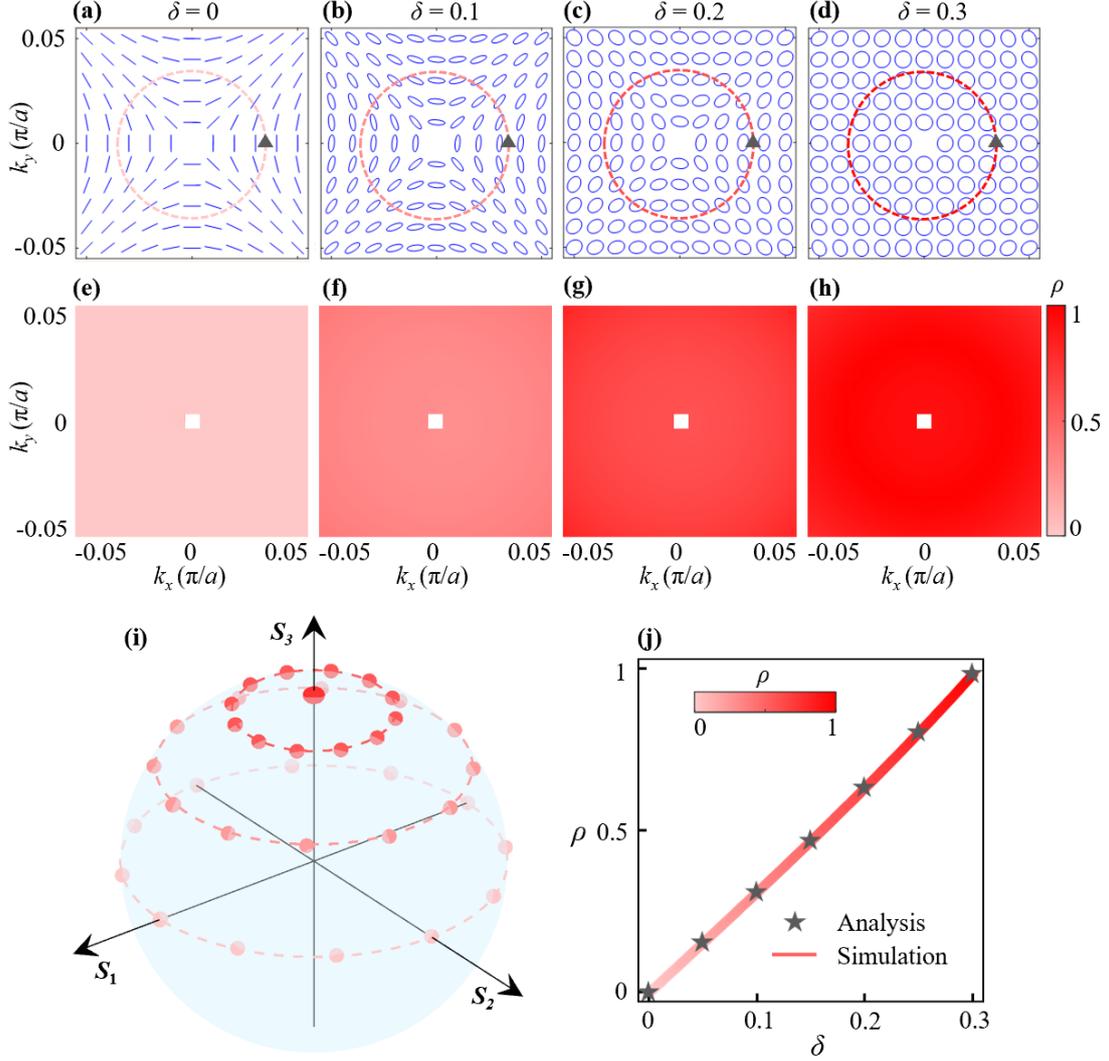

**Figure 2.** Magnetically tunable BICs with arbitrary polarization in the MO PhC slab. (a-d) The evolution of far-field polarization of the MO PhC slab and (e-h) their corresponding ellipticity with increasing external magnetic fields $\delta$ = 0, 0.1, 0.2, and 0.3, respectively. The far-field polarization gradually evolves from linear to circular as $\delta$ increases from 0 to 0.3. (i) The far-field polarization of the iso-frequency contour in momentum space (red dashed circles in (a)-(d)) mapped on the Poincaré sphere with $\delta$ = 0, 0.1, 0.2, and 0.3, respectively. (j) The simulated (red line) and analytical (gray stars) ellipticity of the far-field polarization with different external magnetic field strengths at the gray triangles in (a)-(d) with $(k_x, k_y) = (0.03\pi/a, 0)$.

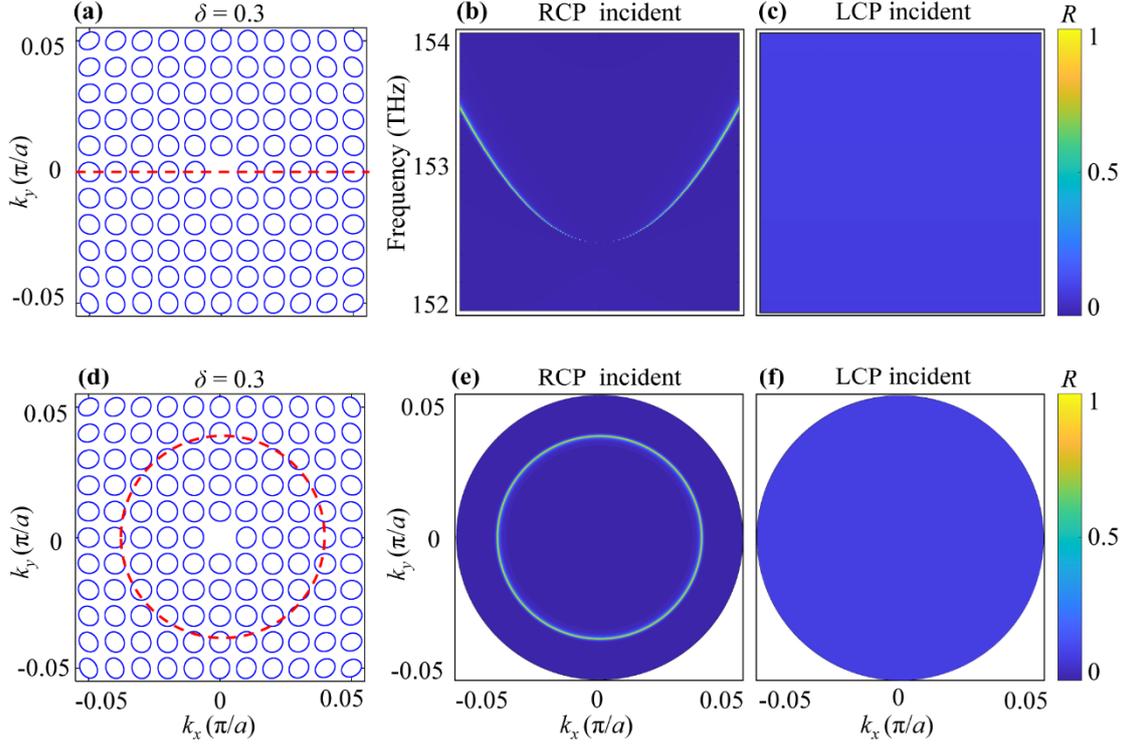

**Figure 3.** Off-Γ chiral emission of the MO PhC slab over a broad frequency range and multiple incident angles. (a) Far-field polarization around the symmetry-protected BICs at the Γ point with $\delta = 0.3$. The red dashed line represents a straight path with fixed $k_y = 0$, and $k_x$ varying from $-0.05\pi/a$ to $0.05\pi/a$. (b, c) Reflection spectra of the MO PhC slab under (b) RCP and (c) LCP incidences along the red dashed line in (a), respectively. (d) The same as (a). The red dashed circle represents the closing path of the iso-frequency contour at $f = 153$ THz in momentum space. Reflection spectra of the MO PhC slab illustrated by (e) RCP and (f) LCP incidences along the red dashed circle in (d), respectively.

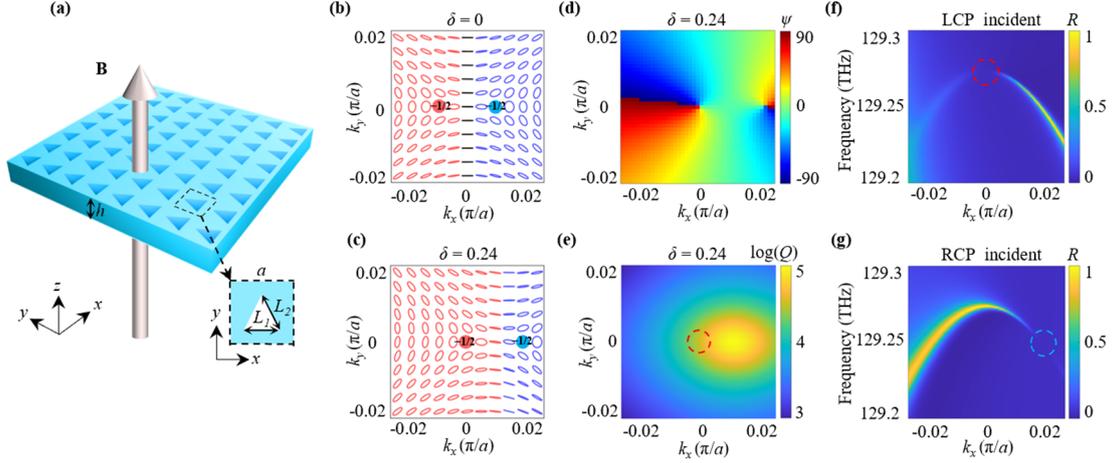

**Figure 4.** At-Γ intrinsic chiral BICs generated by breaking the $C_2$ and TRS symmetries simultaneously. (a) Schematic of a MO PhC slab under perpendicular external magnetic fields. Inset: unit cell of the MO PhC slab. (b-c) Far-field polarization of the MO PhC slab with (b) $\delta = 0$ and (c) $\delta = 0.24$, respectively. When $\delta = 0$, a pair of C points ($q = -1/2$) with opposite chirality are split from the Γ-point BICs by breaking $C_2$ symmetry. When $\delta = 0.24$, an intrinsic chiral BICs is generated by moving the C points to the Γ point. (d-e) Simulated (d) azimuthal angle map and (e) Q factor of the far-field polarization with $\delta = 0.24$. The π-phase change around the Γ point and the high Q factor confirm the existence of an intrinsic chiral BICs at the Γ point. (f-g) Reflection spectra of the MO PhC slab with $\delta = 0.24$ for the (f) LCP and (g) RCP incidences, respectively. The vanishing points of the reflection spectra are indicated by the red and cyan circles.